\RequirePackage{snapshot}

\documentclass[a4paper]{article}
\usepackage{graphicx}

\usepackage{amsmath}
\usepackage[inline]{enumitem}
\usepackage{amsthm}
\usepackage{amsfonts}
\usepackage{wrapfig}
\usepackage{authblk}
\usepackage[dvipsnames]{xcolor}

\newcommand{\degree}{\ensuremath{^\circ}}

\newcommand{\autocite}[1]{\cite{#1}}
\newcommand{\textcite}[1]{\cite[]{#1}}

\begin{document}

\title{Ensemble-Based Well Log Interpretation and Uncertainty Quantification for Geosteering}

\author[1]{Nazanin Jahani}
\author[2]{Joaquin Ambia Garrido}
\author[1]{Sergey Alyaev}
\author[1]{Kristian Fossum}
\author[1]{Erich Suter}
\author[2]{Carlos Torres-Verdín}

\affil[1]{NORCE Norwegian Research Centre, Bergen, Norway}
\affil[2]{The University of Texas at Austin
Hildebrand Department of Petroleum and Geosystems Engineering, 
Austin, USA
}


\maketitle

\begin{abstract}

The costs for drilling offshore wells are high and hydrocarbons are often located in complex reservoir formations. To effectively produce from such reservoirs and reduce costs, optimized well placement in real-time (geosteering) is crucial.
Geosteering is usually assisted by an updated formation evaluation obtained by well-log interpretation while drilling.
A reliable, computationally efficient, and robust workflow to interpret well logs and capture uncertainties in real-time is necessary for this application.
An iterative ensemble-based method, namely the approximate
Levenberg Marquardt form of the Ensemble Randomized Maximum Likelihood (LM-EnRML) is integrated in our formation evaluation workflow.
We estimate model parameters, resistivity and density in addition to boundary locations, and related uncertainties by reducing the statistical misfit between the measurements from the well logging tools and the theoretical measurements from the forward tool simulators.
The results of analyzing several synthetic cases with several types of logs 
verified that the proposed method can give good estimate of model parameters by employing as few as 40 ensemble members and 2--10 iterations. By comparing the CPU time, we conclude that the proposed method has at least about 10--125 times lower computational time compare to a common statistical method, such as Metropolis-Hastings Monte Carlo. In addition, the ensemble-based method can run in parallel on multiple CPUs.
The reliability and speed of well-log interpretation is normally sensitive to several parameters such as the distances between the formation boundaries and the logging tool, model parameter’s contrast, formation layer thickness and well inclination. 
Testing the method on a case inspired from a real field also yielded accurate formation evaluation.
Thus, the proposed ensemble-based method has been proven robust and computationally efficient to estimate petrophysical formation properties, layer boundaries and their uncertainties, indicating that it is suitable for geosteering.

\end{abstract}

\newcommand{\todosaly}[1]{{\color{red} Sergey: #1}}

\section{Introduction}

Real-time geological interpretation and uncertainty quantification of Logging While Drilling (LWD) measurements enable updating of the initial geomodel leading to better geosteering decisions, aiming to increase resource recovery of geothermal or hydrocarbon reservoirs, and reduce the operation cost and drilling risks \cite[]{Alyaev2019a}.
The interpretation is challenging; traditional deterministic and gradient-based minimization inversion methods start from an initial geomodel, and iteratively update it by minimizing the misfit between data from the real measurements (observed data) and the forward simulated measurements (theoretical data).
Such inversion is sensitive to initial guess and regularization, gets stuck in local minima, and does not quantify interpretation uncertainties. Mitigation of these limitations is problem specific and is often only available in proprietary codes \cite[]{baker2014}.

Markov Chain Monte Carlo (MCMC) based methods on the other hand avoid these problems completely.
These methods rely on iteratively solving the forward problem for different proposed solutions, and finally reach an optimal solution, along with an estimation for the uncertainty of the properties. 
The forward solver is typically the most computationally expensive part of this algorithm. 
A comprehensive albeit brute-force approach, like the Metropolis-Hastings method, will require hundreds of thousands of forwards runs to establish a solution \cite[]{metropolis1953equation,bottero2016stochastic}. 
To mitigate this problem, clever ways to reduce the sampling space, without reducing the accuracy of the final solution, have been proposed: \cite{shen2017statistical} implemented a Hamiltonian Monte Carlo (HMC) method, and \cite{SPWLA-UTAustin-2019} proposed a Bayesian inference approach. These approaches could be applied in a geosteering time constrained setting, however they consider the location of the boundaries (between the geological layers) a known input. To address this issue, \cite{veettil2020bayesian} proposed a sequential Monte Carlo filter, which has the extra benefit of not assuming symmetric and Gaussian probability distributions. 
Nonetheless the method assumes constant layer thickness and constant formation dip, which are the geomodel parameters.

Several papers have demonstrated that the ensemble-based methods, such as the ensemble Kalman filter, are robust, and fast, approaches for conditioning high-dimensional state- and parameter-values to data, and to quantify their posterior uncertainties in a satisfying manner \cite[]{evensen1994sequential, aanonsen2009ensemble}. 
\cite{chen2015optimization,luo2015ensemble} applied ensemble-based methods for joint estimation of boundaries and resistivity in a geosteering setting, but were limited to geomodels with two-tree contrasts and synthetic electro-magnetic (EM) measurements.

In this paper we demonstrate that the iterative ensemble smoother
is a preferable real-time interpretation and uncertainty quantification method for across many problems in formation evaluation.
We use a version of it called Levenberg-Marquardt Ensemble Randomized Maximum Likelihood (EnRML, introduced in \cite{Chen2013}) to estimate formation boundaries and petrophysical properties.
We do so across several logging while drilling (LWD) acquisition setups:
The shallow logs, such as neutron density, are utilized to estimate the fine scale reservoir properties of the layers close to the well.
Moreover, extra-deep directional electro-magnetic (Deep EM) measurements,  
which give more information about formation boundaries and reservoir properties for layers further away from the tool \cite[]{antonsen2018geosteering}, are utilized to interpret petrophysical properties on coarser scale \cite[]{baker2014,larsen2015extra}. 
The ensemble-based method transparently uses all prior geological knowledge and updates the associated uncertainties using the acquired data. 

In this work, the reliability, robustness and speed of the proposed ensemble-based method are verified by employing a series of synthetic examples. 
In one of the examples, density is the unknown model variable. In the remaining examples, resistivity and layer boundaries are the  unknown variables. 
Furthermore, we consider a formation model inspired from the Goliat field in the Barents Sea whose variables are resistivity and layer boundaries.
The posterior uncertainty of the estimated model variables is quantified in all studied cases.

The rest of the paper is structured as follows:
Section 'Ensemble-based interpretation' describes our workflow. 
Section 'Model input and output' describes input and output of the forward simulators.
The numerical results for the current study are presented in section 'Numerical Results' and the main conclusions are summarised in Section 'Conclusions'.

\section{Method: Ensemble-based interpretation}

\begin{figure} 
    \begin{center}
    \includegraphics[width=\textwidth]{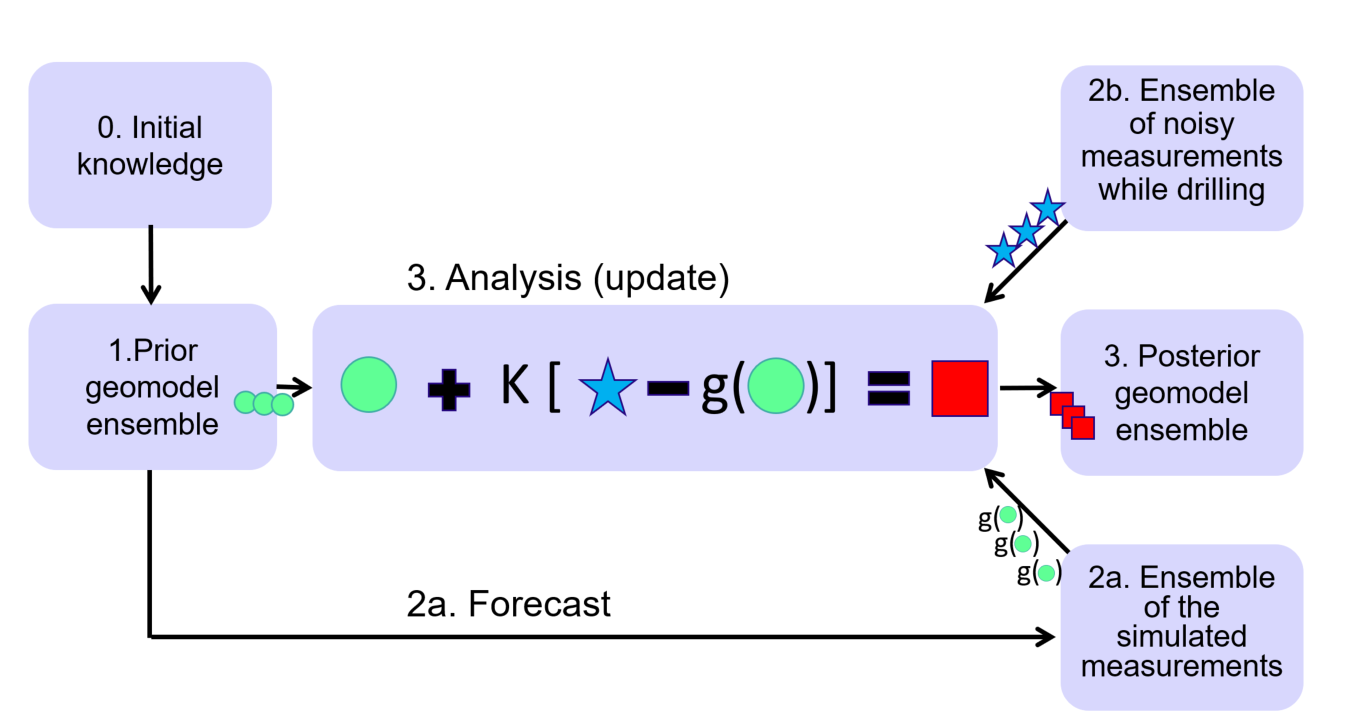}
    \caption{\label{fig:method} Ensemble-based method to estimate formation petrophysical properties and formation boundaries by conditioning the prior earth-model to the LWD measurements.
    The green circles are realizations of the prior model, blue stars are realizations of LWD measurements, and red squares are the output of the analyze step, which is the posterior ensemble.
    More detail about the equation in the analysis step is in the method section.}
    \end{center}
\end{figure}

The diagram in Figure \ref{fig:method} shows the ensemble-based interpretation workflow. 
The method contains three main components: \begin{enumerate*}  \item setting up the prior model, \item performing the forecast step to simulate measurements, the output of this step is theoretical data (2\,a), and including LWD measurements, which are observed data (2\,b),\item and updating the prior model by incorporating the the simulated and observed LWD measurements (analysis) \end{enumerate*}. 
The output of the analysis component is the posterior model.
In the following, we explain the method in more detail.

\subsection{The prior ensemble}

The prior ensemble contains all our prior knowledge, obtained from geological interpretation of the available information.\\
In step 0, we generate 40--100 realizations comprising the prior ensemble $N_e$.
Each realization is a member of the ensemble.

The prior ensemble, $x_n$, is represented by the selected model parameters.
For our study, we chose to include petrophysical properties such as density, resistivity and layer boundaries in every position along the well path and is combined with the given well path to predict the logs.\\
If the prior ensemble is not specified on the well planning stage it can be approximated as a Gaussian or a uniform distribution using the given single geomodel (deterministic base interpretation) as the ensemble mean.
The Gaussian distribution is generated with the given mean model and a specified variance, where the variance represents uncertainties in the model parameters. 
The prior ensemble serves as regularization for the further probabilistic inversion.

\subsection{Incorporating simulated and observed measurements}
In order to update the prior model, the forward simulated measurements, obtained from the forecast step, are compared with the realization of LWD measurements:
\subsubsection{Forecast step and simulated measurements}

In the forecast step,
we use the forward simulator $g(x_n)$ on each of the prior ensemble members to generate the simulated measurements (theoretical data), which is presented in Box 2\,a in Figure \ref{fig:method}.
We present the forward model using in this study in Section: input and output.

\subsubsection{Observed LWD measurements}


LWD measurements Box 2\.b is represented by a vector of real measurements, $d_{j, true}$, perturbed with a vector of additive measurement noise $\epsilon_j$ (see Equation \ref{eq:data}), where $j$ indicates each realization.

The measurement noise is assumed to be uncorrelated Gaussian with zero mean and a specified variance. We define the variance relative to the value of the observed data. More specifically, we let the standard deviation equal a percent value of the observed data. In the numerical study, the standard deviation is defined as 1-5\,\% of the measured data, which in turn is squared to give the variance of the measurement noise. 

\begin{equation}
\label{eq:data}
    d_{j}  =d_{j,true} + \epsilon_j  \quad \text{for } j = 1,\ldots, N_e 
\end{equation}

\subsection{Analysis and update the prior model}

At the analysis step (step 3 in Figure \ref{fig:method}), the prior ensemble $x_{n,j}$ is updated to reduce statistical misfit with the observed measurements.
 
The output of the analysis step is the posterior ensemble, $x_{(n+1),j}$, which is represented in Box 3 in the figure. 
The detail of the analysis step is as follows.

Here, the model variables are conditioned to realizations of the measurements, $d_{j}$ by minimization of the following objective function.

\begin{equation}
    O(x)= \frac{1}{2}\|x - x^{\text{prior}}\|_{C_x} + \frac{1}{2}\|g(x) - d\|_{C_d},
\end{equation}
where
\begin{equation}
    \|\cdot\|_{C} = \left( \cdot \right)^TC^{-1}\left( \cdot \right).
\end{equation}
Formulating the minimization using the Gauss-Newton approach with the Levenberg-Marquardt modification of the Hessian, the iterative update is given by \cite[]{DeanBook}
\begin{align}
    x_{n+1} = & x_{n} - \left[\left(1+ \lambda_n\right)C_x^{-1} + G_n^TC_d^{-1}G_n\right]^{-1}\nonumber \\
    &\times \left[C_x^{-1}\left(x_n - x^{\text{prior}}\right) + G_n^TC_d^{-1}\left(g(x_n) - d\right) \right],
    \label{eq:LM-enrml}
\end{align}
where $G_n$ is the sensitivity matrix.

The EnRML approximate the sensitivity matrix, and the state covariance matrix $C_x$, using information from the ensemble.

To this end, we define the scaled and centred ensemble of states
\begin{equation}
\Delta X_n = \left[x_{(n,0)},\dots, x_{(n,N_e)}\right]\left(I_{N_e} - \frac{1}{N_e}11^T \right)/\sqrt{N_e-1},
\label{eq:state_en_pert}
\end{equation}
and the scaled and centred ensemble of forecast models 
\begin{equation}
    \Delta D_n= \left[g(x_{(n,0)}),\dots, g(x_{(n,N_e)})\right]\left(I_{N_e} - \frac{1}{N_e}11^T \right)/\sqrt{N_e-1}.
    \label{eq:forcast_en_pert}
\end{equation}
The ensemble approximation to the sensitivity matrix is then
\begin{equation}
    G_n = C_d^{1/2}\Delta D_n \Delta X_n^{-1},
    \label{eq:en_sens}
\end{equation}
and the ensemble approximation to the prior-state covariance matrix is
\begin{equation}
    C_x = \Delta X_0 \Delta X_0^T.
    \label{eq:en_cov}
\end{equation}

Following~\cite[]{Chen2013}, we replace $C_x$ in the Hessian term of~\ref{eq:LM-enrml} with $C_{x_{n}}$, and we neglect the terms containing the state mismatch. This gives the approximate LM-EnRML update equation
\begin{equation}
    x_{n+1,j} = x_{n,j} - \left[\left(1+ \lambda_n\right)C_{x_{n}}^{-1} + G_n^TC_d^{-1}G_n\right]^{-1}G_n^TC_d^{-1}\left(g(x_{n,j}) - d_j\right).
\end{equation}
Using the Sherman-Woodbury-Morrison matrix inversion formulas, this is rewritten as
\begin{equation}
    x_{n+1,j} = x_{n,j} -  C_{x_{n}} G_n^T\left[\left(1+ \lambda_n\right)C_d + G_nC_{x_{n}} G_n^T\right]^{-1} \left(g(x_{n,j}) - d_j\right)
\end{equation}

Inserting~\eqref{eq:state_en_pert}-\eqref{eq:en_cov}, and simplifying gives
\begin{equation}
    x_{n+1,j} = x_{n,j} - \Delta X_n \Delta D_n^T\left[\left(1 + \lambda_n\right)I_d + \Delta D_n \Delta D_n^T\right]^{-1}C_d^{-1/2} \left(g(x_{n,j}) - d_j\right).
\end{equation}
To stabilize the method, we perform a truncated Singular Value Decomposition (SVD) of $\Delta D_n$
\begin{equation}
    \Delta D_n = U_p,S_p,V_p^T,
\end{equation}
where the subscript $p$ denotes the truncation. We retain the number of singular values such that their partial sum represents a certain percentage of the sum of all the singular values.
In this numerical study, the singular values are kept such that their sum corresponds to 99\% percent of the sum of all the singular values.
Inserting the SVD gives the iterative scheme utilized in this work
\begin{equation}
    x_{n+1,j} = x_{n,j} - \Delta X_n V_p S_p \left[\left(1+\lambda_n\right)I_p + S_p^2 \right]^{-1}U_p^T C_d^{-1/2} \left(g(x_{n,j}) - d_j\right).
    \label{eq:y}
\end{equation}
Equation~\ref{eq:y} is repeated until convergence. Here, this is defined when the relative improvement in the data misfit is below a threshold. For each iteration that gives a reduction of the objective function, we let $\lambda_{n+1} = \lambda_{n}/10$. In addition, we set the initial value, $\lambda_0$, close to the value of the initial data misfit.

\section{Model input and output } 

AS we described earlier, the core of the our interpretation method is the analysis step, which the forward simulated and observed measurements are its inputs, and the updated geomodel from the inversion formulas is its output. We explain the forward and inverse models in the following.

As we explained in the forward model section, the input for the forward simulator is the geomodel, defined by petrophysical properties and formation boundaries, and the well path:
The geomodel realizations in this work are 1D horizontally layered geomodel that has been extended in the horizontal direction, where all layer boundaries have the same dip.

The number of layers is kept constant.
Layers are characterized by an initial petrophysical properties (resistivity or density) and  their boundaries.  In this work we assume no anisotropy. 

The well path is defined by inclination.
The log was sampled each meter along the measured depth (MD) of the well.
All sampling points are incorporated simultaneously into the ensemble-based model.

\subsection{Forward model}

In this work we apply two separate forward simulators: one for the density log and one for the Deep EM logs. Their output is on the same format as the output of the real logging instruments.

\subsubsection{Density model}
The density simulations are conducted using the forward simulator developed by \cite[]{mendoza2010linear} which relies on flux sensitivity functions. A detailed explanation of the method can be found in the work by \cite[]{luycx2020simulation}.

\subsubsection{Deep EM model}
The Deep EM forward simulator is a deep neural network trained on the output from a commercial simulator software as explained in  \cite{alyaev2020modeling}.
Our forward simulator is trained to respond to up to seven layers (three above and three below the logging tool). The model inputs are the six boundaries of the layers: three above and three below the measuring instrument; the seven resistivities of the layers; and the relative dip between the layers and the instrument.

The schematics of the Deep and Extra Deep EM instruments can be found in \cite{baker2014}.
The DNN model outputs 13 log traces (22 including maximum sensitivity angles) that are typically transmitted in real time \cite[]{alyaev2020modeling}.



\section{Numerical results}

We shall demonstrate that the proposed ensemble-based method is reliable, robust and computationally efficient for interpreting both shallow and deep logs, as well as for quantifying interpretation uncertainties.
For this purpose, we first construct synthetic examples where we estimate layer bulk densities and resistivities using the shallow density and Deep EM logs respectively.
To further study the performance of the proposed ensemble method, we consider a case, inspired from the Goliat field, where the method uses the Deep EM logs to estimate the layer resistivity and boundaries.
We evaluate the computational efficiency of our proposed ensemble method by performing the same example with Metropolis-Hastings Monte Carlo method \cite[]{metropolis1953equation}.

\subsection{Bulk density estimation}

In this example we apply the proposed ensemble-based method to interpret density logs in thinly laminated formations by constructing a synthetic case with layer thicknesses varying between 0.5\,m and 5.5\,m; we assume each layer has uniform density. Layers are horizontal (zero dip angle).
The prior ensemble is a realization of density in each layer is modelled by a Gaussian distribution as we explained in Section of the prior model.
We describe one case with ensemble of 40 and another case with ensemble of 100 realizations; the posterior ensemble mean conveys the estimated density and the posterior standard deviation expresses its uncertainty.
Realization of the observed measurements are generated from the synthetic true measurements by adding a Gaussian relative noise of 1.5\,\%.

\begin{figure} [ht]
\centering
  \includegraphics[width=0.6\textwidth]{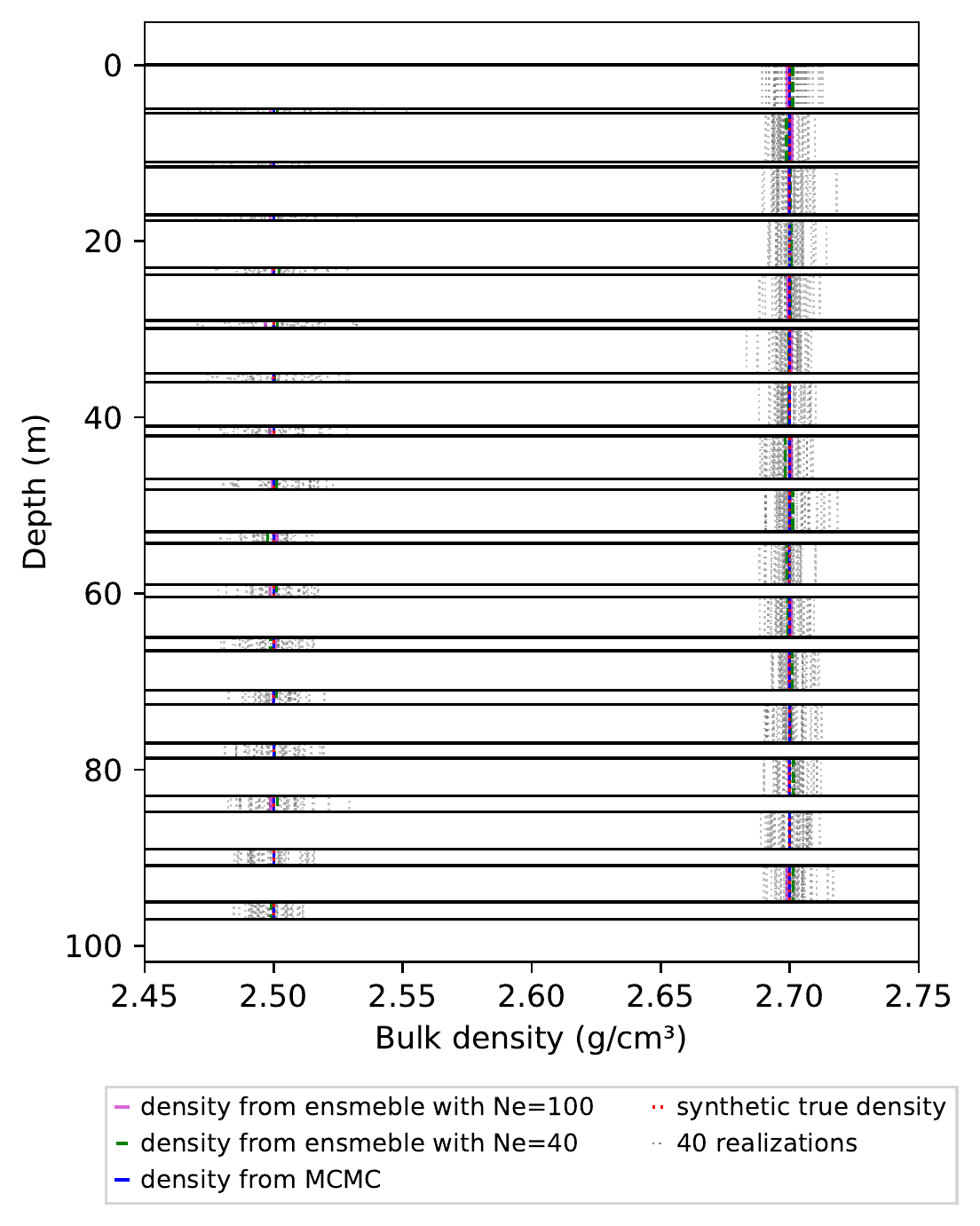}  
   \caption{\label{fig:densityLog} Estimated bulk density ($g/cm^{3}$) and its distribution; true synthetic measurements perturbed with 1.5\,\% noise. Grey lines show each realization for the ensemble with 40 realizations.
   }
\end{figure}

\begin{figure}
\centering
\includegraphics[width=0.5\textwidth]{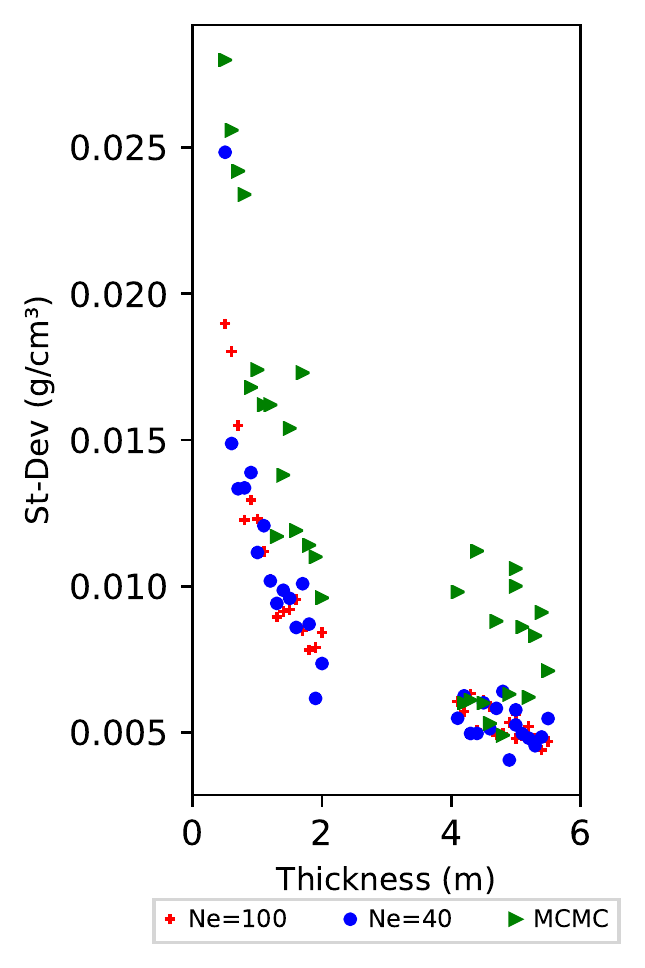}  
\caption{\label{fig:std-density}Posterior standard deviation of bulk density ($g/cm^{3}$) versus layer thicknesses for cases with 40 realizations, 100 realizations and results from MCMC with 10000 sampling steps, 
assuming 1.5\,\% Gaussian noise. There is two group of layer thicknesses: thinner than 2\,m and thicker than 4\,m. As layer thicknesses decrease uncertainty increase.}
\end{figure}

The grey lines in Figure \ref{fig:densityLog} is the posterior realizations. The figure compares estimated density (posterior ensemble mean) from the ensemble with 40 realizations and with the density from the ensemble with 100 realizations in addition to comparing them with the estimated density from MCMC method and with the true density.
It shows that for all the layers, true density (red dotted lines) lies inside of the posterior ensemble (grey lines).
They also agree well with the results from MCMC (dotted blue lines) with 10000 sampling steps. 
In this example, three iterations of Equation \ref{eq:y} of 
LM-EnRML method are sufficient.
The posterior ensemble means are very close in cases with 40 and 100 realizations, thus we conclude that ensemble of 40 is sufficient. 
This means that the total number of forward runs is as low as $3 \times 40 = 120$, compare to 10000 needed for the MCMC.
Figure \ref{fig:std-density} shows that by decreasing layer thicknesses, the value of the posterior standard deviation increases; as expected because of the higher relative impact of shoulder bed effects \cite[]{SPWLA-UTAustin-2019}.

\subsection{Resistivity estimation}
In this example, we verify the speed and robustness of the proposed method for interpreting the Deep EM logs to estimate layer resistivities and boundaries
of a layer-cake geomodel  as described in section.
For this reason, we study the sensitivity of the method for four parameters: the distance from the logging tool to the layer boundaries, property contrast across the layer boundaries, and layer thicknesses by constructing four synthetic scenarios:
\begin{enumerate}
    \item The logging tool distances to the layer boundaries are varying. The well angle is 80\degree--82.5\degree. The tool is in the layer with low resistivity (1.4\,ohm-m) and is approaching the layer with the highest resistivity (99\,ohm-m). We focus on quantifying uncertainty of boundaries location, which are farther away from the logging tool.
  
    \item Layers with higher resistivity contrast (3\,ohm-m and 50\,ohm-m), equal thicknesses (20\,m). The well path lands with 80\degree\ , and its inclination increases with depth and becomes near-horizontal. We focus on quantifying uncertainty in layers with high resistivity.
    \item Variable layer thicknesses (0.7\,m to 10\,m), constant 80\degree well angle and no resistivity contrast. We focus on quantifying uncertainty in thinner layers.
\end{enumerate}

 Furthermore, in the Goliat field example section we consider an example inspired from the Goliat field in the Barents Sea. In this example we compare the estimated resistivity, recovered from the ensemble-based method, with the one from Metropolis Hastings Monte Carlo method.
 
The prior models in all examples are constructed with 40 or 100 realizations with a Gaussian distribution.
Logs (observed data) are perturbed with 1--5\,\% relative noise and sampled along the measured depth of the well every one meter. 

\subsubsection{Case 1: Decreasing distance from the logging tool to the target layer}

\begin{figure}
    \centering
    \includegraphics [width=0.8\textwidth]{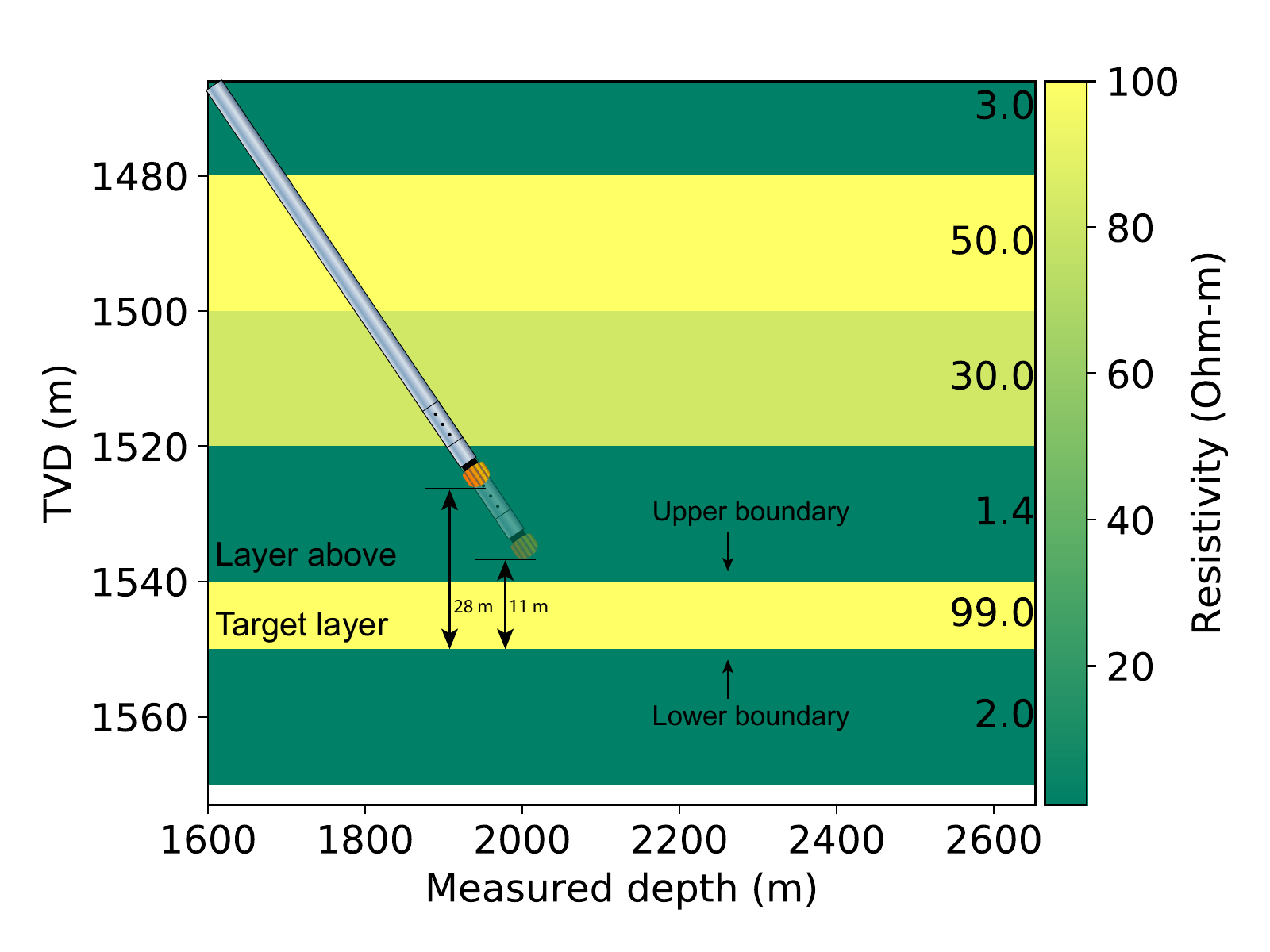}
    \caption{\label{fig:distance-earth}Synthetic case 1: the tool gradually approaches the target layer, sand layer with resistivity of 99\,ohm-m, from the layer above, shale layer, with low resistivity of 1.4\,ohm-m.
    This study considers the starting sampling position at TVD at 1522\,m and the last sampling point at TVD 1539\,m. 
    Thickness of the target layer is 10\,m. TVD of the upper boundary is 1540\,m and the lower boundary is 1550\,m }
\end{figure}

In Figure \ref{fig:distance-earth} we drill towards the target layer, with high-resistivity (99\,ohm-m) and 10\,m thickness, passing through the layer above, low-resistivity layer (1.4\,ohm-m) with 20\,m thickness.
We let 'upper boundary' always denote the top boundary of the target layer, and 'lower boundary' always denote the bottom boundary of the target layer.
The tool distance to the upper boundary decreases gradually from 17\,m to 1\,m: the starting position is at TVD 1522\,m.  and ending position is at TVD 1539\,m.
Standard deviation of the measurement ensemble is 1\% .The number of realization in the ensemble is 100. The standard deviations of the prior ensemble for boundaries is 2\,m.

\begin{figure}
    \includegraphics[width=0.5\textwidth]{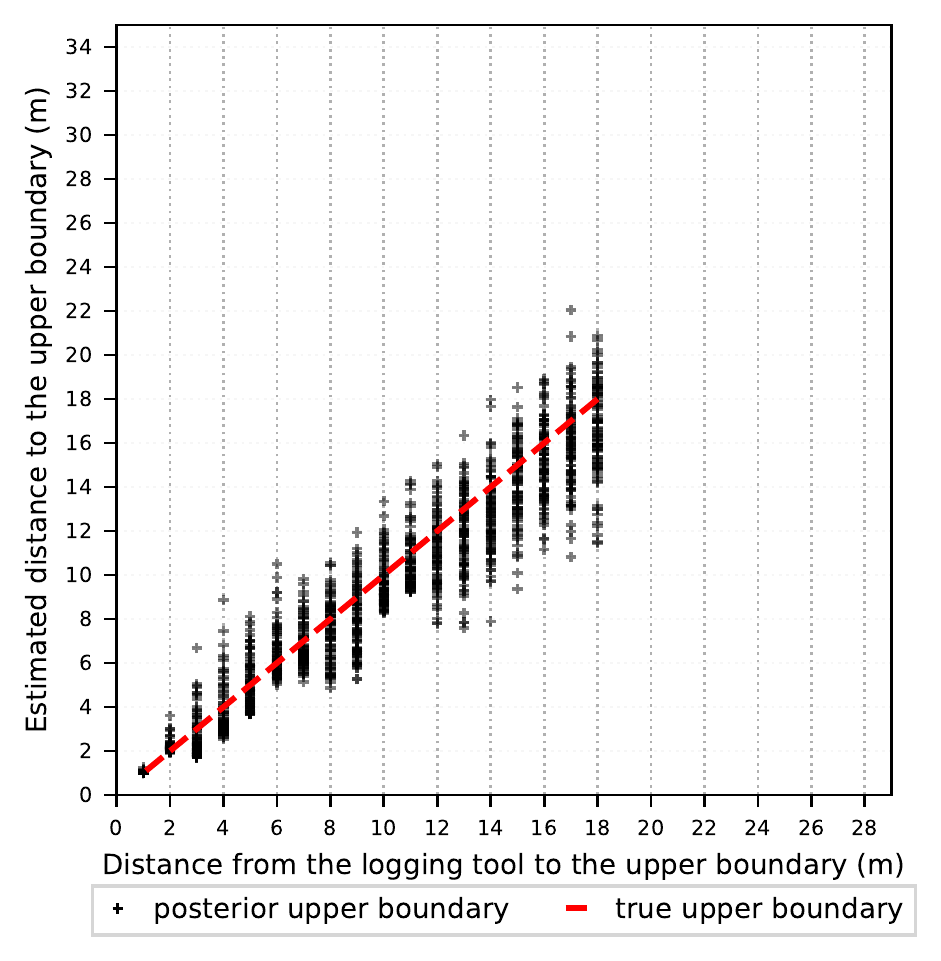}
    \includegraphics[width=0.5\textwidth]{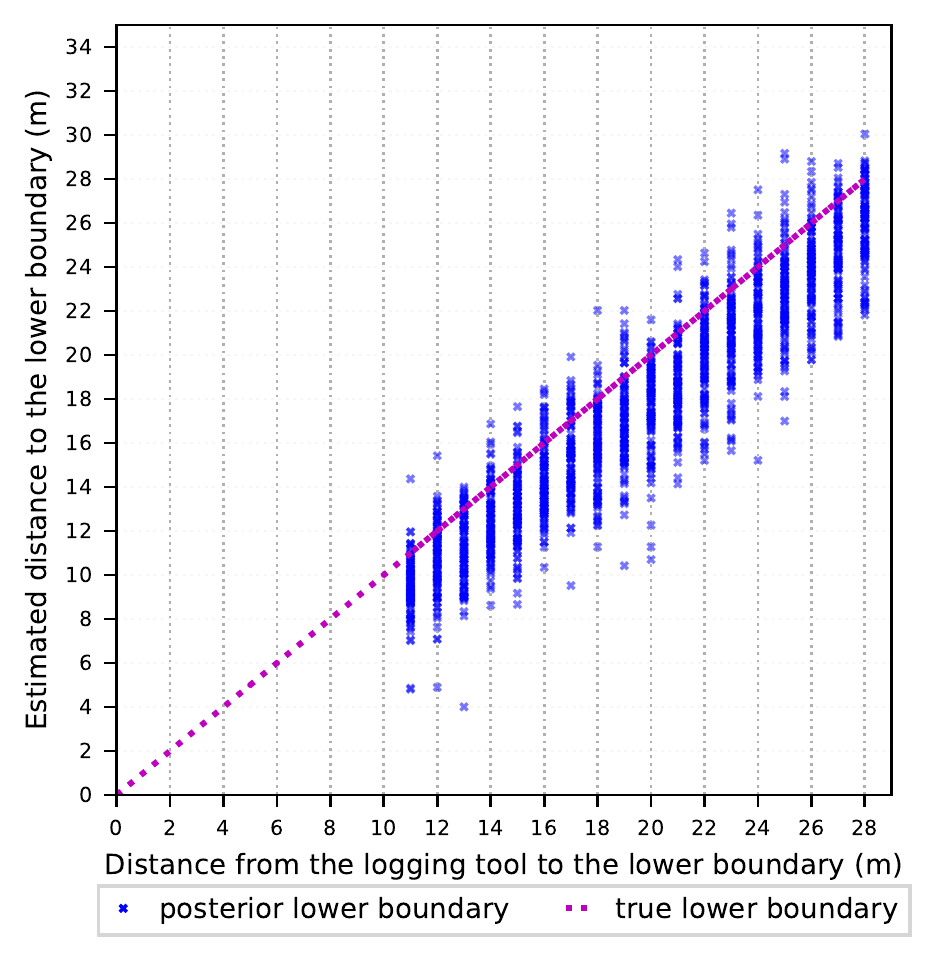}
    \caption{  \label{fig:Distance-two-layers}Synthetic case 1: the two dotted and dashed lines illustrate the true upper (left-side) and lower boundaries (right-side) of the target layer. The target layer and its boundaries are visualized in figure \ref{fig:distance-earth}.
    From top to bottom along y-axis of this figure, the tool distance to the target layer boundaries decreases; as the tool is approaching the target layer, the estimation of the boundary position become more certain.
    }
\end{figure}

\begin{figure}[ht]
\centering
\includegraphics[width=1\textwidth]{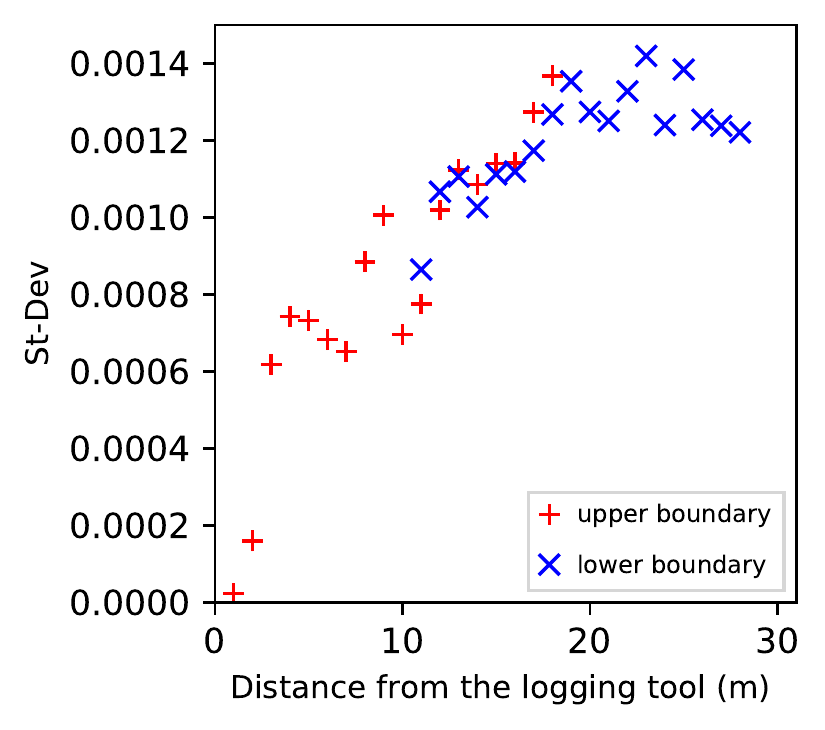}
\caption{\label{fig:Dis-std-tvd}Synthetic case 1: posterior fractional standard deviation of the estimated target layer boundaries versus distance between the boundary and the logging tool. The true value of the upper boundary is 1540\,m and the lower one is 1550\,m.
}
\end{figure}

Figure \ref{fig:Distance-two-layers} shows the posterior distribution of the estimated upper and lower boundaries versus distance of the boundary to the logging tool.
We observe as the tool is approaching the target layer, the estimation of the boundary position become more certain.
The results of the proposed ensemble-based method show that the estimation of the distance to the nearest boundary (upper boundary) are robust and accurate, with very low uncertainty, at the distance less than 2\,m, and the true boundary is recovered by the data.
At the distance from 3\,m--11\,m, the boundary is estimated with less uncertainty in comparison to farther distances: 12\,m--17\,m. 
Not that the well angle increases from 80\degree to 82.5 \degree as the tool approaches to the target layer; The higher uncertainty in some sampling points, closer to the target layer, in comparison to the neighbouring farther points may be due to the effect of varying the well angle. 
The lower boundary is estimated at the range of 29\,m to 11\,m farther away from the tool despite the noise.
The uncertainty decreases as the tool distance become less than 17\,m.
Note, that around 17\,m to 29\,m farther from the lower boundary, the prediction stays robust but with increased uncertainty due to weaker signal.
The true boundaries are covered by data (Figure \ref{fig:Distance-two-layers}).
Figure \ref{fig:Dis-std-tvd} shows the standard deviation of the estimated target layer boundaries versus the tool distance to the boundaries.
We observe expected high uncertainty in the position, as the logging tool is farther away from the boundaries.
The higher uncertainty in the layer location of the boundary is preserved in the layer farther away from the tool.
This is in agreement with the result presented by \textcite{larsen2019detecting}, where the tool detects the position of the upper boundary with low uncertainty in the distance around 13.5--14.8 m and the position of the lower boundary with higher uncertainty at around 16.2-18.2\,m, and the tool is able to detect the boundary with uncertainty as the logging tool is approaching the layer at around 33-40\,m.\\


\clearpage
\subsubsection{Case 2: Formations with high resistivity contrast between neighbouring layers}
\begin{figure}[ht]
\centering
\includegraphics[width=0.7\textwidth]{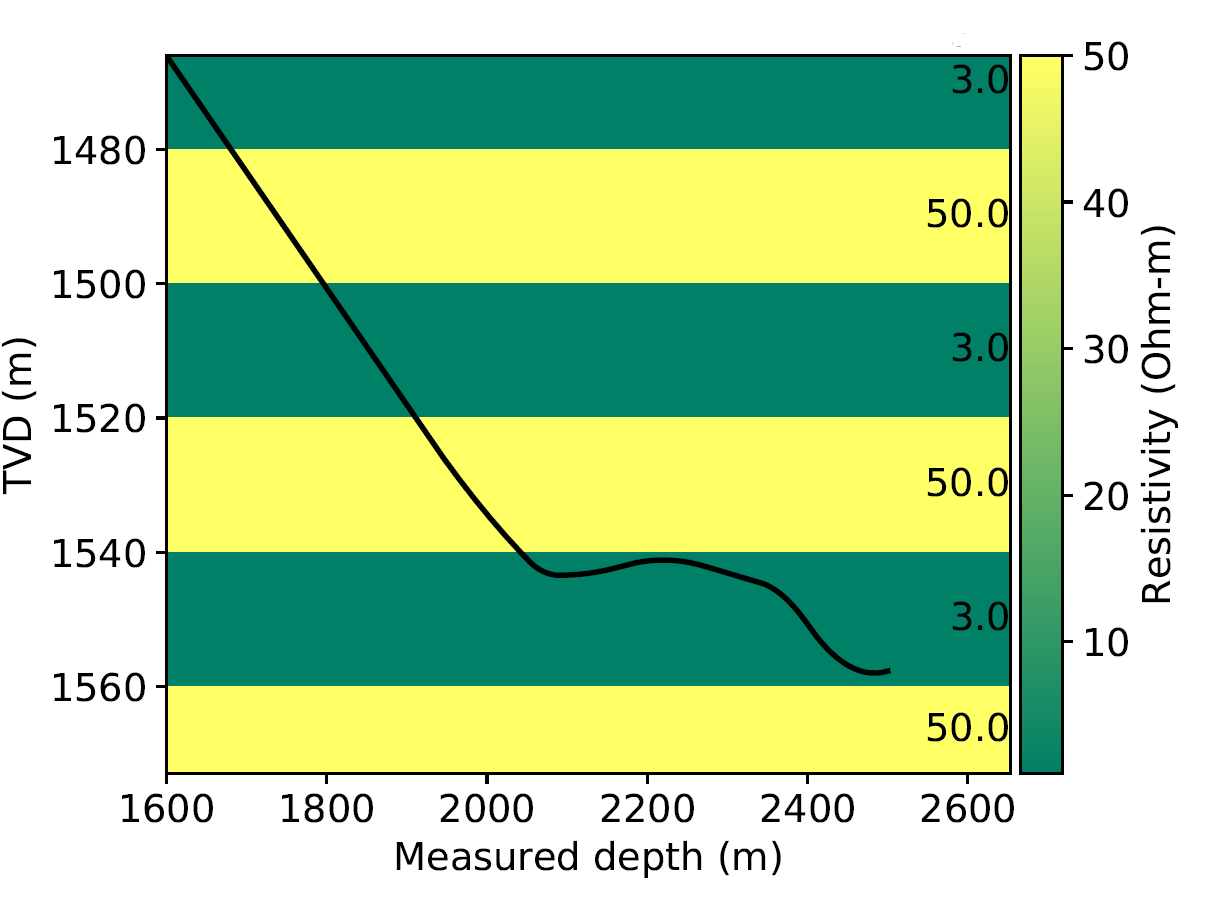}
\caption{\label{fig:Well-contrast}Synthetic case 2: 1D horizontally layered  geomodel that has been extended in the horizontal direction with equal layer thicknesses and high resistivity contrast. Five layers from the top are drilled with high angle ( 80 to 90\degree ). The logging tool does not reach to the 6\textsuperscript{th} layer.}
\end{figure}

The geomodel for this case, Figure \ref{fig:Well-contrast}, contains 6 horizontal layers with equal layer thicknesses, with high resistivity contrast between neighbouring layers (3\,ohm-m and 50\,ohm-m). Five layers from the top are drilled with high to near-horizontal angle.
The last layer is not penetrated. 

\begin{figure}[ht]
 \centering
  \includegraphics[width=0.5\linewidth]{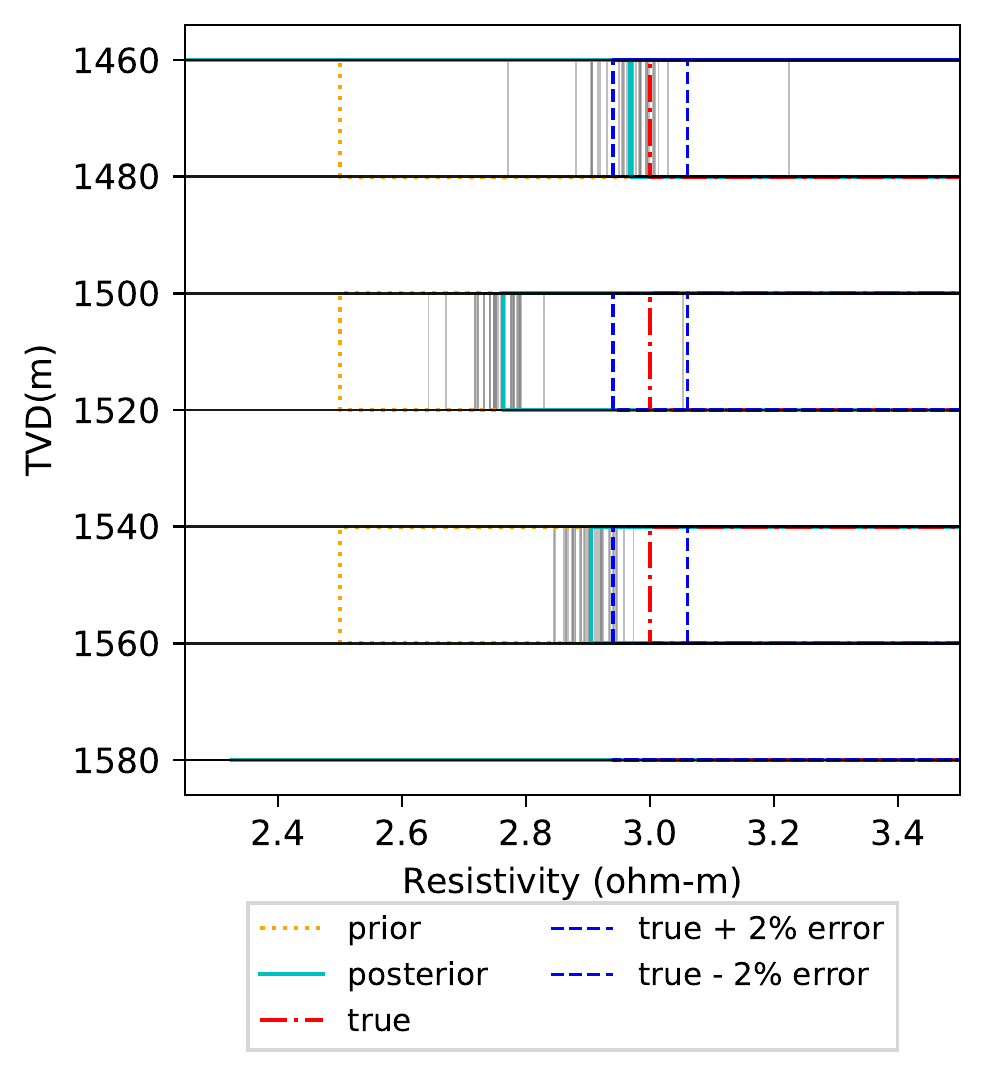} 
 \hfill
  \includegraphics[width=0.5\textwidth]{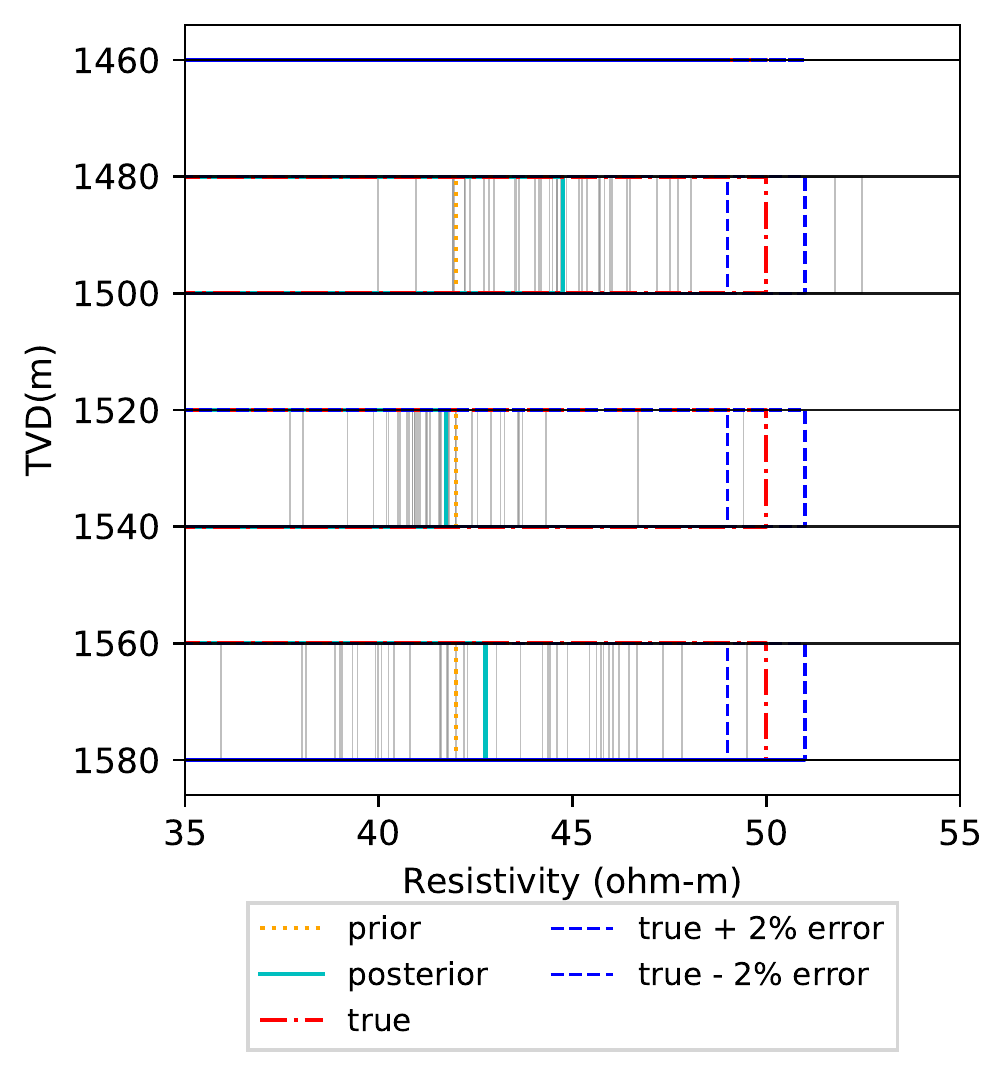} 
  \caption{\label{fig:res-con}Synthetic case 2: Estimated layer resistivities (mean of the posterior ensemble) in comparison with the true resistivity and the prior mean resistivity for layer with resistivity of 3 (left-side) and 50\,ohm-m (right-side). 
  The measurement noises are 2\% relative error.}
\end{figure}

The plot in figure \ref{fig:res-con} displays the prior and estimated posterior resistivity for each layer after 5 iterations of LM-EnRML method.
The number of realizations is 40 hence the number of forward runs is 200 for this case.
The true resistivity is covered by the posterior ensemble in the first layer. Figure \ref{fig:res-con} shows ensemble of the posterior distribution is more spread in the layers with the higher resistivity, indicating the higher uncertainty in these layers. The highest uncertainty is preserved in the last layer, which is farther away from the tool.
 
 
\begin{figure}[ht]
  \centering
  \includegraphics[width=1\linewidth]{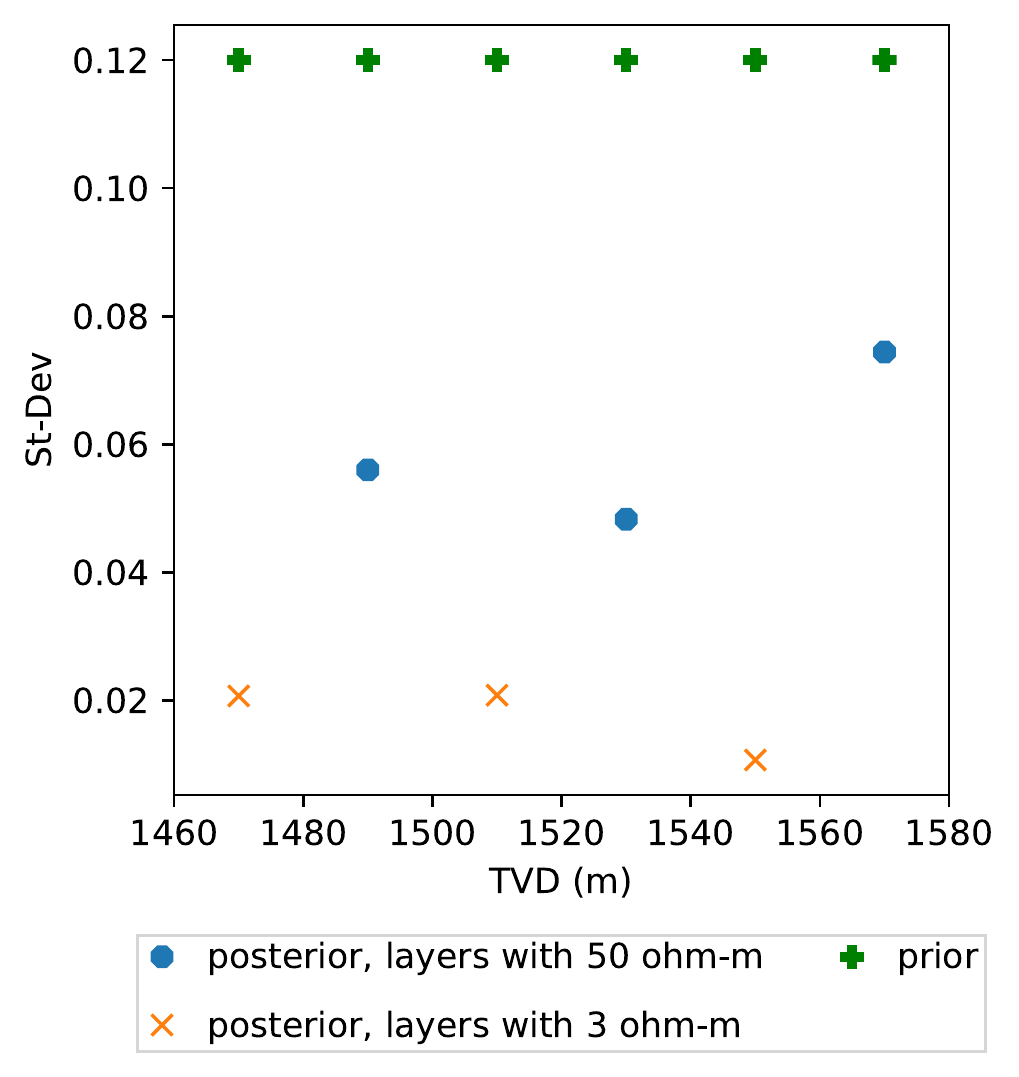}  
  \caption{ \label{fig:contrast-std}Synthetic case 2: posterior standard deviation, divided by layer's mean estimated resistivity: higher posterior standard deviation for layers with resistivity 50\,ohm-m.}
\end{figure}

Figure \ref{fig:contrast-std} compares the standard deviation of the prior ensemble withe the posterior ensemble. It represents that the uncertainty decreases after the update.
Figure \ref{fig:contrast-std} indicates, that layers with higher resistivity (50\,ohm-m in this case) have higher posterior standard deviation, so higher uncertainty in comparison to the layers with lower resistivity (3\,ohm-m in this case). 
The very low standard deviation in the layer (1540-1560) is due to the high number of measurements, which are along the well path, in this layer.
The last layer, farther away from the logging tool, presents visibly higher standard deviation.
There are no measurements for the last bottom layer (1560--1580 m TVD). This demonstrates the ability of the ensemble method to preserve the uncertainty in the regions of the model where the amount of data is insufficient. 

\clearpage
\subsubsection{Case 3: Formations with varying thicknesses}

In order to verify the applicability of the method to estimate resistivities of thinner layers, and to study the effect of layer thicknesses on uncertainty,  we construct a synthetic case with 100 realizations with varying layer thicknesses from 0.7\,m to 10\,m.
In order to focus only on the effect of layer thicknesses on sensitivity, we assume no resistivity contrast between neighbouring layers and that the true resistivity is 50\,ohm-m.
The mean resistivity of the initial prior model is 45\,ohm-m. Measurements error is 3\%, applied as standard devaluation to the true measurements.
The well path inclination is 80\degree. 

\begin{figure}[ht] 
\centering
  \includegraphics[width=1\textwidth]{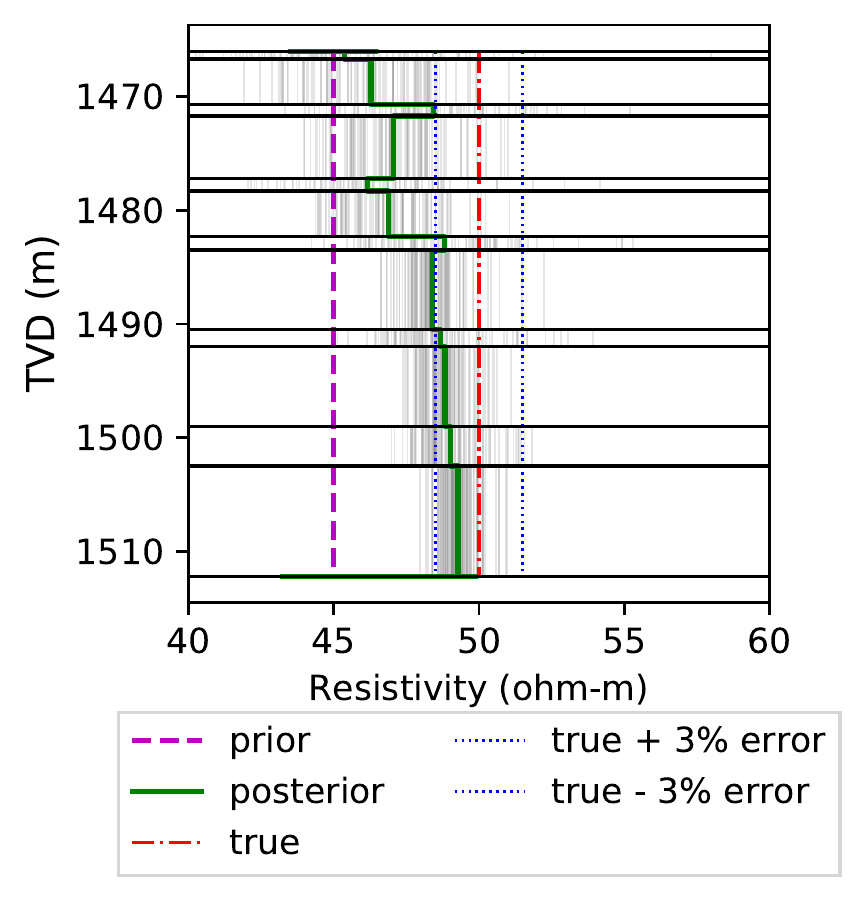}  
  \caption{\label{fig:tickness-res}Synthetic case 3: estimated resistivity in cases with varying thickness (0.7\,m -- 10\,m); the thinner layers have higher uncertainty. 
  In the figure, 'posterior' indicates the mean of all realizations in the posterior ensemble.
    Grey lines show posterior realizations.
   }
\end{figure}

\begin{figure}[ht]
  \centering
  \includegraphics[width=1\textwidth]{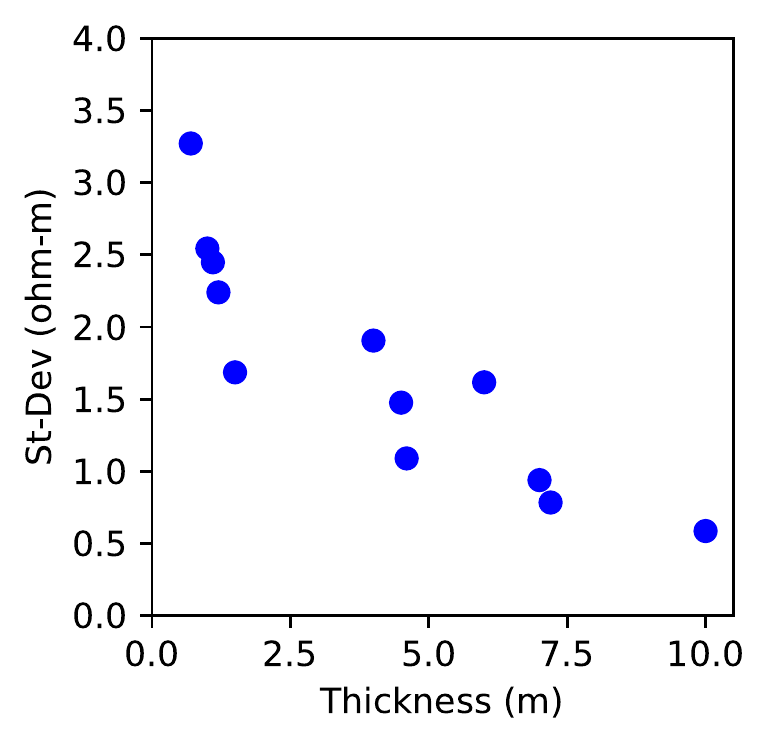}  
  \caption{\label{fig:tickness-std}Synthetic case 3: posterior standard deviation of the layer resistivities (\,ohm-m) versus layer thickness: thinner layers have higher standard deviation.}
\end{figure}

Figure \ref{fig:tickness-res} shows the distribution of the posterior resistivity (grey lines) and its mean (solid blue lines). 
We see the thinner layers have noticeably higher uncertainty in the posterior ensemble. 
Inspecting the standard deviations of the layer resistivities (Figure \ref{fig:tickness-std}) we see the same relation: there is higher standard deviation for the resistivities of thinner layers.
This is expected because of the higher relative impact of shoulder bed effects.
These effects also make the problem more non-linear and its solution requires 4 to 6 LM-EnRML iterations.

\clearpage
\subsection{The Goliat field example} 

\begin{figure}[ht]
  \begin{center}
    \includegraphics[width=0.7\textwidth]{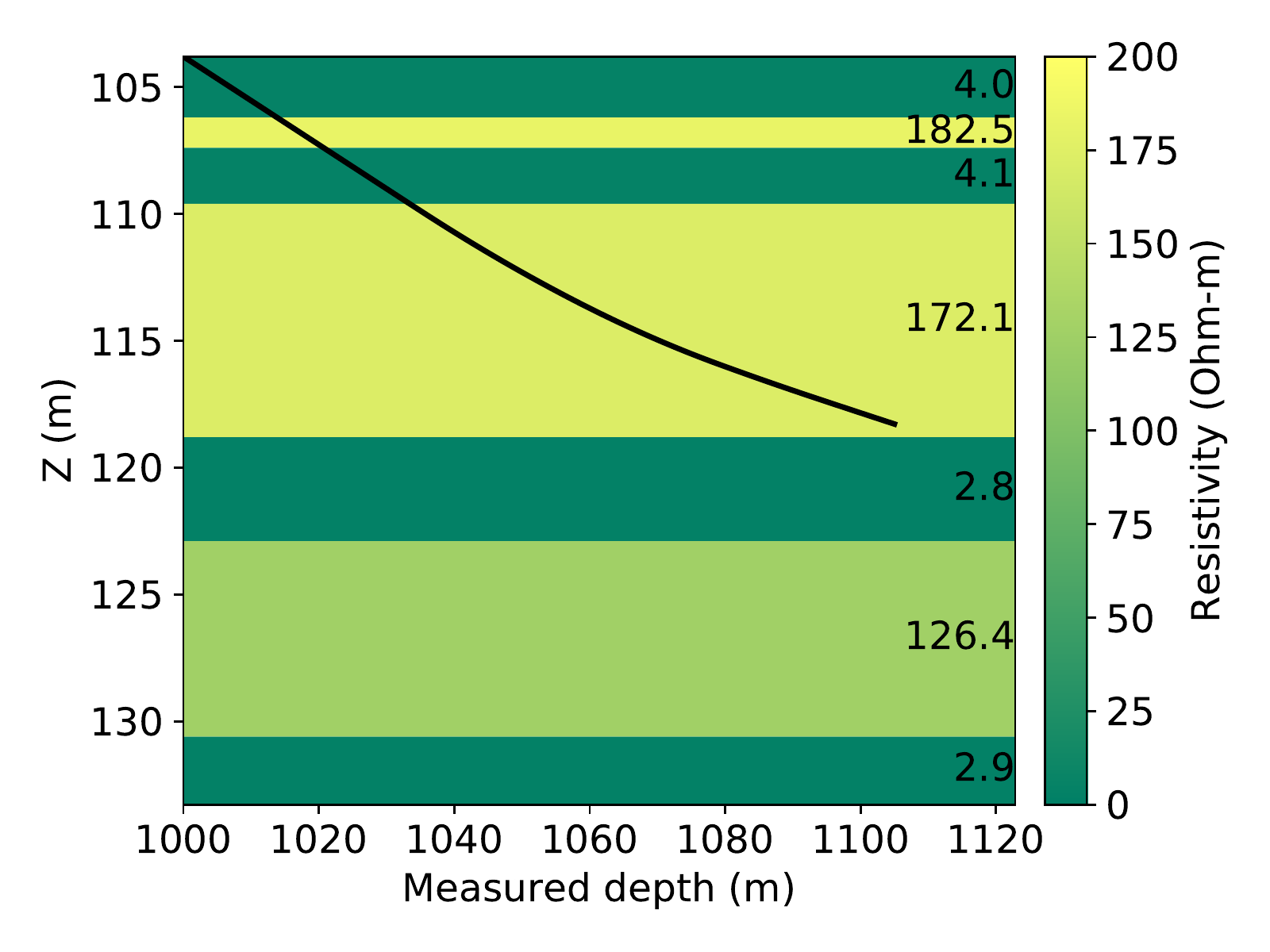}
  \end{center}
  \caption{\label{fig:wellA}A 2D section with a local coordinate system that shows the approximation of the geology near the well-landing point for 'Well' A from \textcite{larsen2015extra} . Both the earth model and the well path have been rotated 5\degree\ to make the layering horizontal;  here, MD 1000 corresponds to x000 in \textcite{larsen2015extra}. The well inclination increases with depth (from 80\degree to 85\degree).
 }
\end{figure}

We further demonstrate the speed and applicability of the proposed method by testing it on a case inspired by a real geosteering operation in the Goliat field in the Barents Sea, described in \textcite{larsen2015extra}.
The geomodel covers a small unfaulted section just before the well ('Well A') lands in the drilling target in the Upper Kobbe formation.
This formation is characterized by thin layers and high resistivity contrasts. The aim of this example is to get an indication of the behaviour of the method in a realistic geosteering situation. 
Figure \ref{fig:wellA} shows a simplified 2D section inspired from \textcite{larsen2015extra}, which was used for testing of the forward model in \textcite{alyaev2020modeling}.
Model variables are the layer resistivities and boundaries, which are estimated simultaneously.
We first demonstrate resistivity estimation results and compare the results with resistivity estimation from Metropolis-Hastings Monte Carlo, afterwards we show the results of the joint estimation of the boundaries.

\subsubsection{Resistivity estimation}
\begin{figure} [ht]
   \centering
   \includegraphics[width=1\textwidth]{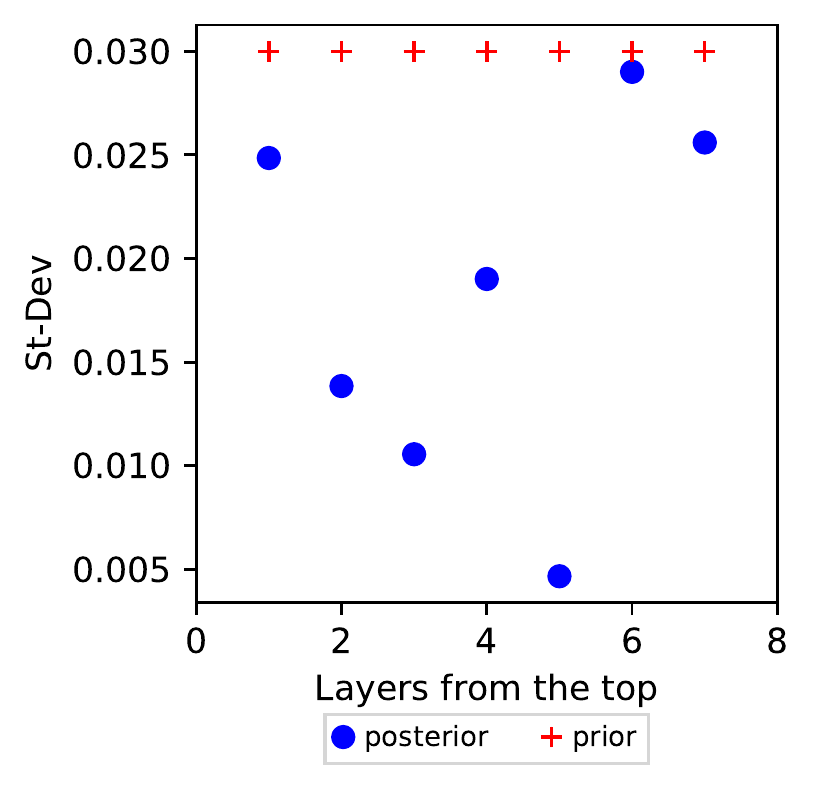}
   \caption{\label{fig:wellA-std-res} Posterior fractional standard deviations of the layer resistivities in comparison with the initial errors in the prior ensemble: High uncertainties in the last two layers farther away from the tool are preserved by the model.}
\end{figure}

\begin{figure}
    \centering
    \includegraphics[width=1\textwidth]{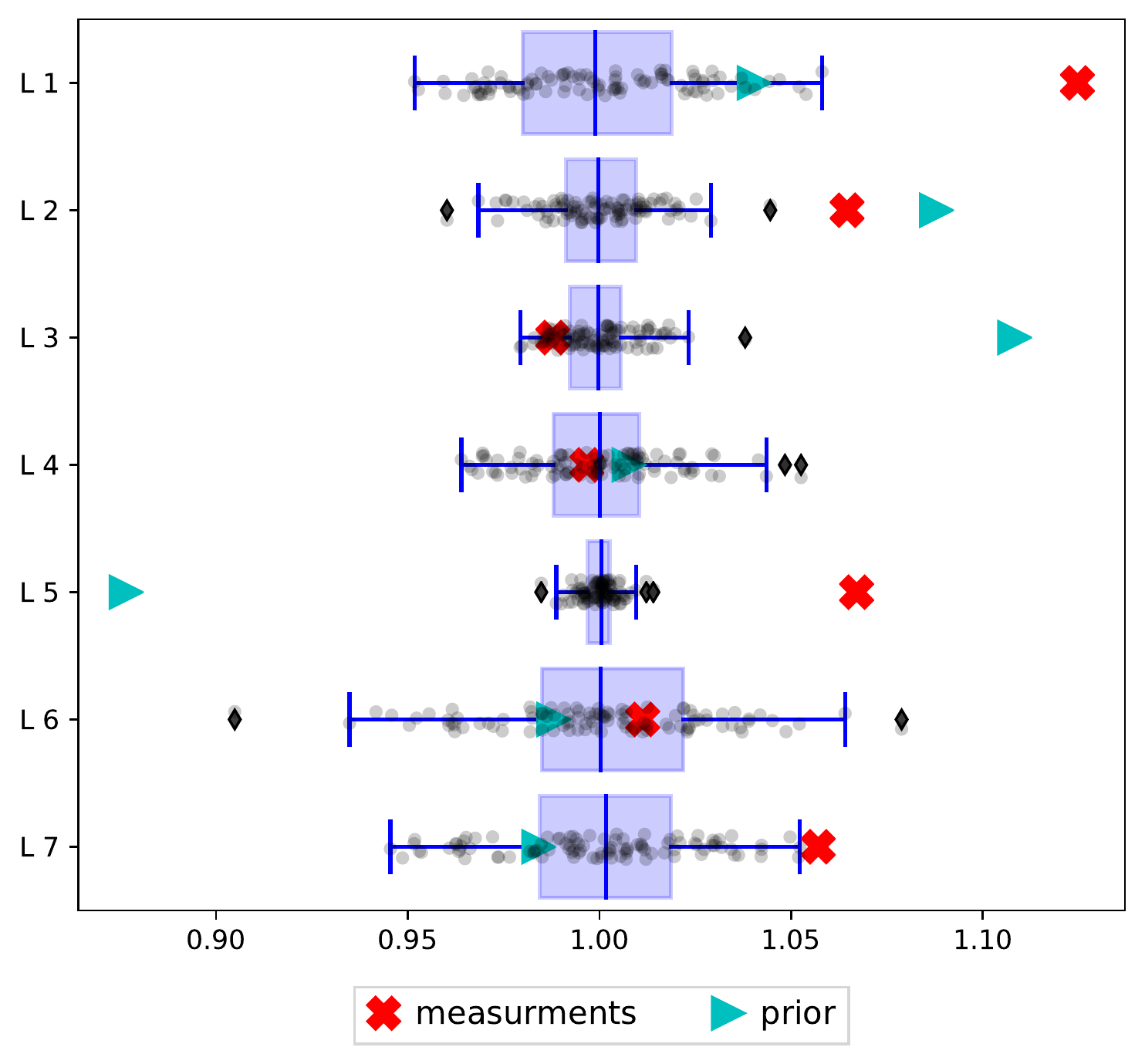}
    \caption{The box-plots visualize the posterior resistivity distribution, divided by its mean value. The mean value of the measurements and the prior ensemble, divided by the mean of posterior, are included in this figure.}
    \label{fig:Boxplot_WellA}
\end{figure}

\begin{figure} [ht]
\centering
\includegraphics[width=0.7\textwidth]{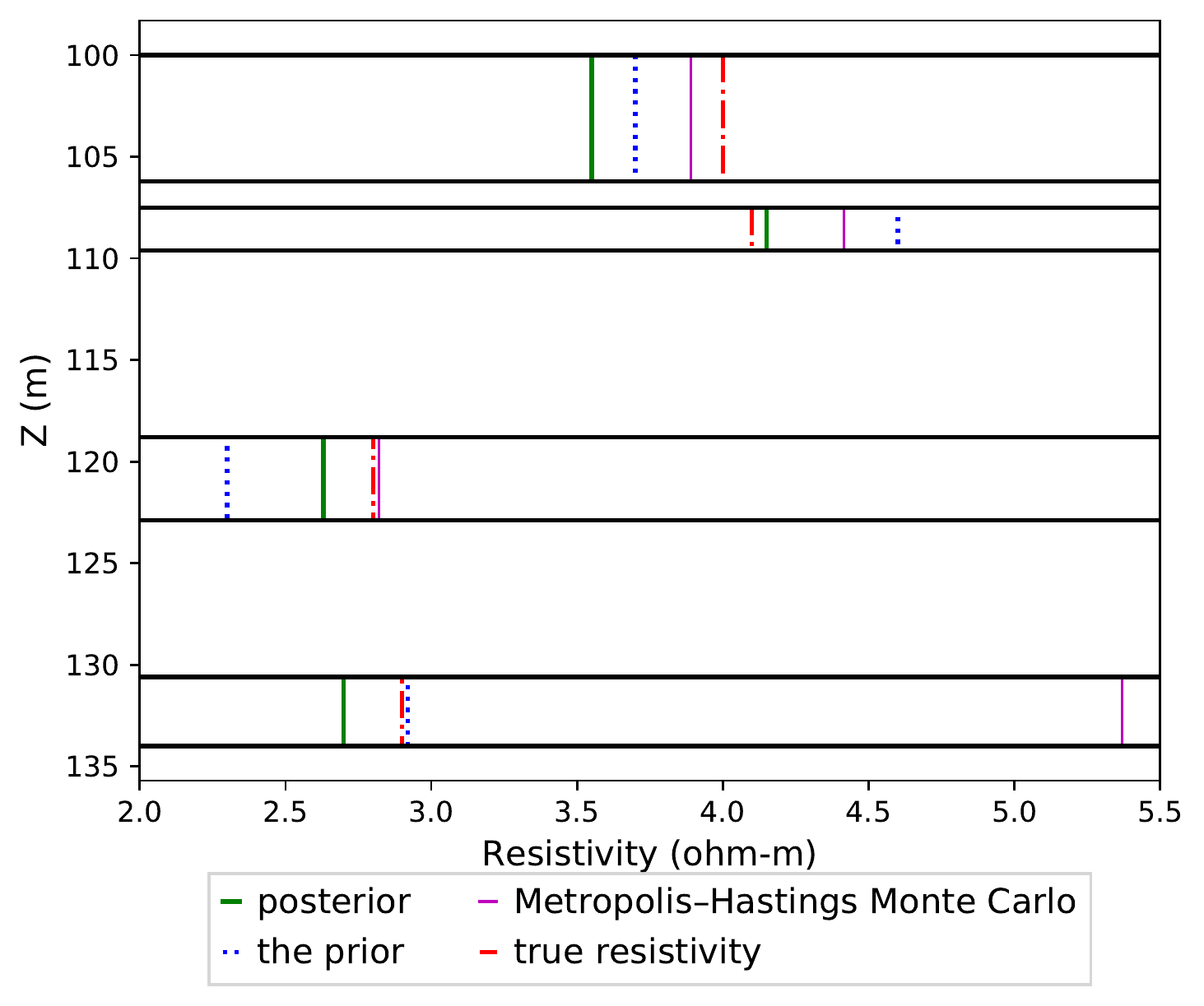}
\hfill
\includegraphics[width=0.7\textwidth]{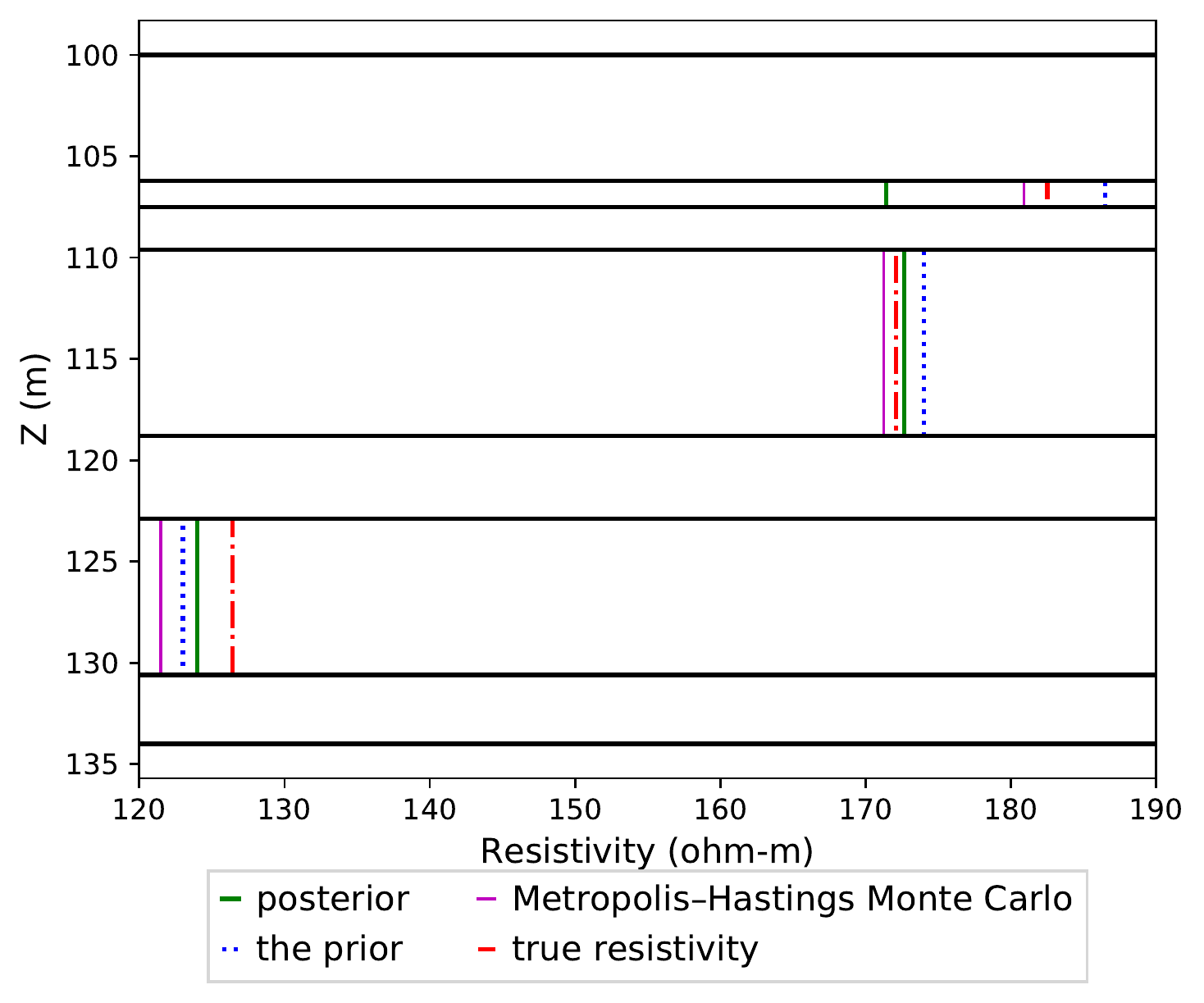}
  \caption{\label{fig:Res-wellA-MC}Comparing the resistivities from the ensemble-based method after 5 iterations with those from Metropolis-Hastings Monte Carlo method after 10000 iterations.}
\end{figure}

The well path is from MD 1000 to MD 1105, and a total of 106 positions along the well path are sampled.
Similar to the synthetic examples in this work, we derive the theoretical measurements for each sample point from the DNN approximation presented in \textcite{alyaev2020modeling}.
We construct the ensemble of the measurements with 1\% standard deviation.
The prior ensemble is set up with 100 realizations. We define the initial prior ensemble with the mean value and the standard deviation. In this model the fraction standard deviation of the prior ensemble is 0.03 (3\% error) for all layer's resistivity, and 0.0025 (0.25\% error) for the boundary locations. 
Figure \ref{fig:wellA-std-res} shows the standard deviation of the posterior resistivities.
This presents the uncertainties decrease after the update, however the model preserves the higher uncertainty of the last two bottom layers, which are farther away from the tool.
The first layer also show higher uncertainty; It is because in this model the well path starts from the local depth of 103\,m, but we chose the depth of the first layer at the local depth of 100\,m,  so there is less measurement points for this layer.
The box-plots in Figure \ref{fig:Boxplot_WellA} visualize the posterior resistivity distribution at each layer separately.
The measurements' mean of the layers 3,4 and 6 lie inside the interquartile range of the box-plot.
The estimated resistivity at Layer 5, which is the first layer below the tool, is between the prior and the measurements. The uncertainty in this layer, which has the lowest resistivity among layers, is the lowest; as we observed in our case 2, the prediction of resistivities in layers with low resistivities is more certain.
It shows the model can estimate the model parameters ahead of the logging tool with better accuracy than its initial estimation. We also perform the simulation with several farther and closer priors to the true value and observe the similar trend.
According to our observations from the previous section, high error in resistivity estimation of the layer 2 can be explained as the layer 2 is a thin layer and has the highest resistivity among layers.
Figure \ref{fig:Res-wellA-MC} compares the estimated resistivity from the ensemble method with the resistivity obtained using the Metropolis-Hastings Monte Carlo method \cite[]{metropolis1953equation} and with the true resistivity. 
The estimated resistivities using Metropolis-Hastings Monte Carlo show higher accuracy for the first two top layers, however layers 3 and 4 in which the data are sufficient, both methods show a comparable accuracy. Both methods estimate the resistivity of the farthest layer from the tool with high error.
The error in the last layers is due to little logging tool sensitivity and can be potentially mitigated by increasing the size of the ensemble or by introducing localization to prevent updates due to the noise.
The ensemble-based method requires 3--5 iterations, yielding 300--500 calls of the forward simulators, while Metropolis-Hastings Monte Carlo requires to check at least 10000 states to minimize the misfit and estimate the resistivity with a comparable accuracy.


\clearpage
\subsubsection{Boundary estimation}
\begin{figure}[ht]
\centering
\includegraphics[width=1\linewidth]{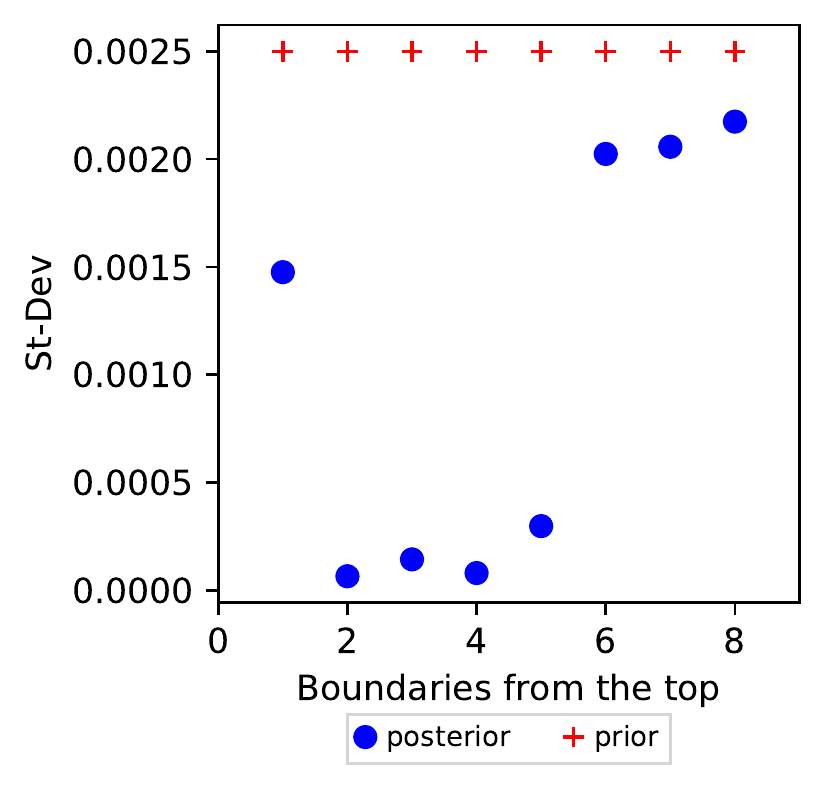}
\caption{\label{fig:wellA-boundary-std} Posterior fractional standard deviation of the layer boundaries of the estimated boundary; higher uncertainties are preserved for the layers farther away from the logging tool.}
\end{figure}

The ensemble-method can update multiple variables simultaneously. 
In this example we estimate the boundaries location for the Goliat field example. We presented its geomodel earlier in Figure \ref{fig:wellA}. The thickness of the layers are not constant, leading to limitation of selecting a prior model with higher variance for the boundaries location.
For this example we use a prior model of the layers with a Gaussian distribution, and
with 25\,cm to 32\,cm standard deviations (0.0025 fractional standard deviation) increasing from the top to the last bottom boundaries.
According to the posterior standard deviation in figure \ref{fig:wellA-boundary-std}, apart from the boundaries far from the logging tool (the first and the last three boundaries), the uncertainties of the estimated boundaries are less than 0.05\%, resulting in near-deterministic inversion despite the modelled uncertainty in the resistivities.
However, higher uncertainties are preserved for the layers farther away from the logging tool.



\clearpage

\section{Conclusion}

With this study we demonstrated the capabilities of the iterative ensemble-based method to estimate petrophysical properties. 
We estimated layer boundaries as well as formation properties such as density and resistivity 
using nuclear density and extra deep logs respectively.
The method is reducing the statistical misfit between the observed LWD measurements and the theoretical LWD measurements obtained from forward simulators and thus estimates the mean and quantifies the uncertainty with posterior distributions.
Even a small improvement in the estimation of layer boundaries may lead to a significant risk reduction when placing the well in an optimal position close to the boundary of a target layer.

First and foremost we verify the method on deep EM measurements.
When the positions of the boundaries are assumed known, 
our interpretation is qualitatively similar to earlier work: estimation is less certain in geological configurations with thin layers and high contrast between each layer's petrophysical properties \cite{SPWLA-UTAustin-2019}.
Furthermore, the presented results indicate that the method is capable of reducing boundary and property uncertainties within the expected range of sensitivities of the modeled deep EM tool: with low uncertainties when the boundary is within 15\,m from the tool \cite[]{larsen2019detecting}.
Finally we verify our method on a case mimicking a historical operation in the Goliat field in the Barents Sea \cite{larsen2015extra}. 
The proposed method not only recovers the mean position of the boundary from noisy measurements, but also provides uncertainty estimates at no extra computational cost.

In absolute terms, the amount of needed forward simulations is at least twenty times lower compare to an MCMC method which gives similar uncertainty quantification. Moreover for our method most of simulations can be executed in-parallel. These qualities make it really attractive for real-time interpretation. 

Even though the ensemble method might not be necessary for inversion of simpler logs, such as density, our implementation of the method for nuclear density opens future possibility for joint inversion of multiple logs in the same framework.
Thereafter, the framework can be used as part of a future highly automatic real-time workflow for geosteering decision support.




\section{Acknowledgements}

Funding:  This work was supported by the research project 'Geosteering for IOR' (NFR-Petromaks2 project no. 268122) which is funded by the Research Council of Norway, Aker BP, Equinor, V{\aa}r Energi and Baker Hughes Norway.

\bibliographystyle{seg}  
\bibliography{bibJabref}

\begin{thebibliography}{}
\itemsep0pt

\bibitem[Aanonsen et~al., 2009]{aanonsen2009ensemble}
Aanonsen, S.~I., G. Nævdal, D.~S. Oliver, A.~C. Reynolds, and B. Vallès,
  2009, The ensemble kalman filter in reservoir engineering: a review: Spe
  Journal, {\bf 14}, 393--412.

\bibitem[Alyaev et~al., 2020]{alyaev2020modeling}
Alyaev, S., M. Shahriari, D. Pardo, A.~J. Omella, D. Larsen, N. Jahani, and E.
  Suter,  2020, Modeling extra-deep em logs using a deep neural network: arXiv
  preprint arXiv:2005.08919, Accepted to Geophysics March-April 2021.

\bibitem[Alyaev et~al., 2019]{Alyaev2019a}
Alyaev, S., E. Suter, R.~B. Bratvold, A. Hong, X. Luo, and K. Fossum,  2019, {A
  decision support system for multi-target geosteering}: Journal of Petroleum
  Science and Engineering, {\bf 183}, 106381.

\bibitem[Antonsen et~al., 2018]{antonsen2018geosteering}
Antonsen, F., M.~E. Teixeira De~Oliveira, S.~A. Petersen, R.~W. Metcalfe, K.
  Hermanrud, M.~V. Constable, C.~T. Boyle, H.~E. Eliassen, D. Salim, and J.
  Seydoux,  2018, Geosteering in complex mature fields through integration of
  3d multi-scale lwd-data, geomodels, surface and time-lapse seismic: Presented
  at the SPWLA 59th Annual Logging Symposium, Society of Petrophysicists and
  Well-Log Analysts.

\bibitem[Bottero et~al., 2016]{bottero2016stochastic}
Bottero, A., A. Gesret, T. Romary, M. Noble, and C. Maisons,  2016, Stochastic
  seismic tomography by interacting markov chains: Geophysical Journal
  International, {\bf 207}, 374--392.

\bibitem[Chen et~al., 2015]{chen2015optimization}
Chen, Y., R.~J. Lorentzen, and E.~H. Vefring,  2015, Optimization of well
  trajectory under uncertainty for proactive geosteering: SPE Journal, {\bf
  20}, 368--383.

\bibitem[Chen and Oliver, 2013]{Chen2013}
Chen, Y., and D.~S. Oliver,  2013, {Levenberg–Marquardt forms of the
  iterative ensemble smoother for efficient history matching and uncertainty
  quantification}: Comput. Geosci.

\bibitem[Deng et~al., 2019]{SPWLA-UTAustin-2019}
Deng, T., J. Ambía, and C. Torres-Verdín,  2019, Fast bayesian inversion
  method for the generalized petrophysical and compositional interpretation of
  multiple well logs with uncertainty quantification: Presented at the SPWLA
  60th Annual Logging Symposium.

\bibitem[Evensen, 1994]{evensen1994sequential}
Evensen, G.,  1994, Sequential data assimilation with a nonlinear
  quasi-geo\-stroph\-ic model using monte carlo methods to forecast error
  statistics: Journal of Geophysical Research: Oceans, {\bf 99}, 10143--10162.

\bibitem[Larsen et~al., 2019]{larsen2019detecting}
Larsen, D., J. Xavier, S. Martakov, J. Gripp, R. Cremonini, A. Pierre, A.
  Hinkle, N. Tropin, M. Sviridov, and T. Lagarigue,  2019, Detecting ahead of
  bit in a vertical well: First EAGE Workshop on Pre-Salt Reservoir: from
  Exploration to Production, European Association of Geoscientists \&
  Engineers, 1--5.

\bibitem[Larsen et~al., 2015]{larsen2015extra}
Larsen, D.~S., A. Hartmann, P. Luxey, S. Martakov, J. Skillings, G. Tosi, and
  L. Zappalorto,  2015, Extra-deep azimuthal resistivity for enhanced reservoir
  navigation in a complex reservoir in the {Barents Sea}: Presented at the SPE
  Annual Technical Conference and Exhibition, Society of Petroleum Engineers.

\bibitem[Luo et~al., 2015]{luo2015ensemble}
Luo, X., P. Eliasson, S. Alyaev, A. Romdhane, E. Suter, E. Querendez, and E.
  Vefring,  2015, An ensemble-based framework for proactive geosteering:
  Presented at the SPWLA 56th Annual Logging Symposium, Society of
  Petrophysicists and Well-Log Analysts.

\bibitem[Luycx et~al., 2020]{luycx2020simulation}
Luycx, M., M. Bennis, C. Torres-Verdín, and W. Preeg,  2020, Simulation of
  borehole nuclear measurements: A practical tutorial guide for implementation
  of monte carlo methods and approximations based on flux sensitivity
  functions: Petrophysics, {\bf 61}, 4--36.

\bibitem[Mendoza et~al., 2010]{mendoza2010linear}
Mendoza, A., C. Torres-Verdín, and B. Preeg,  2010, Linear iterative
  refinement method for the rapid simulation of borehole nuclear measurements:
  Part 2—high-angle and horizontal wells: Geophysics, {\bf 75}, E79--E90.

\bibitem[Metropolis et~al., 1953]{metropolis1953equation}
Metropolis, N., A.~W. Rosenbluth, M.~N. Rosenbluth, A.~H. Teller, and E.
  Teller,  1953, Equation of state calculations by fast computing machines: The
  journal of chemical physics, {\bf 21}, 1087--1092.

\bibitem[Oliver et~al., 2008]{DeanBook}
Oliver, D.~S., A.~C. Reynolds, and N. Liu,  2008, Inverse theory for petroleum
  reservoir characterization and history matching: Cambridge University Press.

\bibitem[Shen et~al., 2017]{shen2017statistical}
Shen, Q., H. Lu, X. Wu, X. Fu, J. Chen, Z. Han, and Y. Huang,  2017,
  Statistical geosteering inversion by hamiltonian dynamics monte carlo method,
  {\it in} SEG Technical Program Expanded Abstracts 2017: Society of
  Exploration Geophysicists,  900--904.

\bibitem[Sviridov et~al., 2014]{baker2014}
Sviridov, M.~V., A. Mosin, Y. Antonov, M. Nikitenko, S. Martakov, and M.
  Rabinovich,  2014, New software for processing of lwd extradeep resistivity
  and azimuthal resistivity data: Society of Petroleum Engineers, {\bf 17}.

\bibitem[Veettil and Clark, 2020]{veettil2020bayesian}
Veettil, D. R.~A., and K. Clark,  2020, Bayesian geosteering using sequential
  monte carlo methods: Petrophysics, {\bf 61}, 99--111.

\end{thebibliography}

\end{document}